# Consistent analytical solution for the response of a nanoscale circuit to a mode-locked laser

Mark J. Hagmann and Logan D. Gibb
NewPath Research L.L.C., 2880 S. Main Street, Suite 214, Salt Lake City, Utah USA 84115
(Dated July 2, 2020)

## ABSTRACT

It is now common practice to solve the Schrödinger equation to estimate the tunneling current between two electrodes at specified potentials, or the transmission through a potential barrier by assuming that there is an incident, reflected, and transmitted wave. However, these two approaches may not be appropriate for applications with nanoscale circuits. A new approach is required because the electron man-free path may be as long as 68.2 nm in metals so it is possible that the wavefunction may be coherent throughout a nanoscale circuit. We define several algorithms for determining the eigenvalues with different sets of the circuit parameters, thus demonstrating the existence of consistent solutions for nanoscale circuits. We also present another algorithm that is being applied to determine the full solution for nanoscale circuits. All of this is done using only analytical solutions of the Schrödinger equation.

## I. INTRODUCTION

In 1991 Kalotas and Lee [1] introduced the transfer-matrix method for solving the one-dimensional Schrödinger equation to model quantum tunneling in arbitrary static potential barriers. Later Grossel, Vigoureux and Baida [2] compared the stability of this method with that of WKB. We were the first to apply the transfer-matrix method with a time-dependent potential when studying the effects of the barrier traversal time in laser-assisted scanning tunneling microscopy (STM) [3]. These simulations guided our development of microwave oscillators based on laser-assisted field emission [4] and the generation of microwave frequency combs by focusing a mode-locked laser on the tunneling junction of an STM [5]. Now we are studying a variant of STM where extremely low-power ($\approx$ 3 atto-watt) microwave harmonics of the laser pulse repetition rate have a signal-to-noise ratio of 20 dB. Extremely low-noise measurements are possible because the quality factor (Q) at each harmonic is approximately $10^{12}$, which is five times that for a cryogenic microwave cavity [6]. Thus, the effect of white noise is considerably reduced. These harmonics can increase the speed and stability in the feedback control of the tip-sample separation in an STM without exposing the sample to the continuous intense static field of $\approx 10^9$ V/m that causes band-bending in semiconductors [7]. The present analysis is essential because we are developing a macroscopic instrument that is coupled to a nanoscale circuit for scanning tunneling microscopy.

## II. DELIMITATIONS

We model electrons tunneling between two ideal metal electrodes with the same work function and zero resistivity but do not consider the effects of superconductivity. The distribution of electron energies and the effects of images of the tunneling electrons at the two electrodes are neglected to obtain relatively-simple analytical solutions. The Dirac equation should be used to include the properties of the electron (Per-Olov Löwdin, personal communication, 1998) but we still follow the convention of using the Schrödinger equation.

## III. CRITERIA FOR CONSISTENT SIMULATIONS OF NANOSCALE CIRCUITS

### 1. Coherent propagation of the wavefunction

Others have solved the Schrödinger equation for the transmission through static potential barriers with three or more sections at different constant potentials where there is an incident and a reflected wave at one end and a transmitted wave at the other end [8], [9], [10], [11], [12]. This approach has pedagogical value, but it does not consider how the incident, reflected, and transmitted waves may interact at the voltage source ("battery") and the leads so it may not be appropriate for modeling a nanoscale circuit. When modeling a scanning tunneling microscope generally quantum effects are only considered in the region between the tip and the sample [13].

Now we consider the possibility that quantum effects may occur throughout a nanoscale circuit because the electron mean-free path is as long as 68.2 nm in metallic elements [14]. Surprisingly, the effective resistance of a nanoscale wire is proportional to the mean free path of the electrons [15]. This effect has been described and analyzed [16]. However, in a nanoscale circuit the short path length would limit the effects of this resistance. Now we address nanoscale devices with quantum tunneling that already introduces a high electrical resistance into the circuit so the effects of the increased resistance may not be problematic in our application.

### 2. Circuit models for resistive loads and voltage sources

In quantum simulations generally a voltage source is not shown in the diagram but it is implied by the specified the potentials. Now, in allowing for the coherent transfer of the wavefunction throughout a nanoscale circuit, a voltage source is represented by a jump in the potential at a specific location or a linear rise in the potential over a specific distance. Figure 1 shows how a linear variation of the voltage, diagrammed as a linear variation of the potential, may be used to represent a voltage source or a load resistor having a specific length, and lossless connections are represented by horizontal lines. Note that in this diagram the energy is greater than the potential to avoid quantum tunneling.

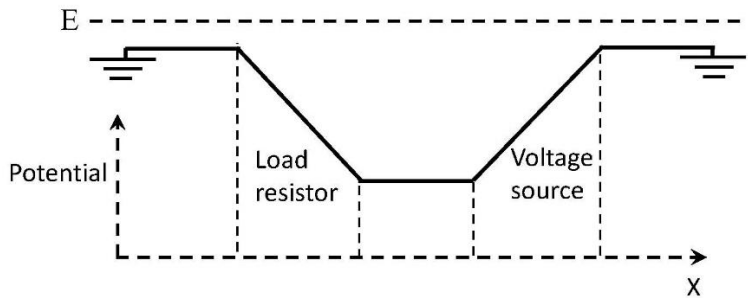

Fig. 1. Symmetric model for a load resistor, a voltage source, and lossless connectors.

In a one-dimensional static problem, the electrical current density in the x-direction is given by Eq. (1) as the product of the probability current density and the charge of the electron. The effective value of the resistance for a resistor model may be estimated by dividing the voltage drop across the simulated resistor by the product of the electrical current density and the effective cross-sectional area of the resistor. It is possible to model a resistor as a lumped circuit element by having a sharp drop in the potential but we may also taper the potential as shown in Fig. 1 for a resistor with a specified length. Adding resistors to a model is the converse to adding voltage sources, and horizontal lines, showing a constant potential, represent wires having no significant resistivity.

$$J_X(x) = \frac{-ie\hbar}{2m}\left(\psi \frac{d\psi^*}{dx} - \psi^* \frac{d\psi}{dx}\right) \quad (1)$$

Figure 2 is a simplified closed-loop model of a nanoscale circuit with a load resistor and a tunneling junction. While linear or "jump" models are not shown for either the load resistor or the voltage source in this figure, it is possible to approach the problem in the following manner: First, the effective voltage is specified which is defined as the voltage from the voltage source minus the drop on the load resistor. The parameters ϕ, $U_0$, a, E, and S may be specified so the wavefunction may be determined. Then the electrical current density may be determined using Eq. (1). An effective cross-sectional area for the resistor may be input for the user to estimate the resistance. Then the voltage drop on the load resistor may be calculated so that the full potential of the voltage source may be determined to complete the solution.

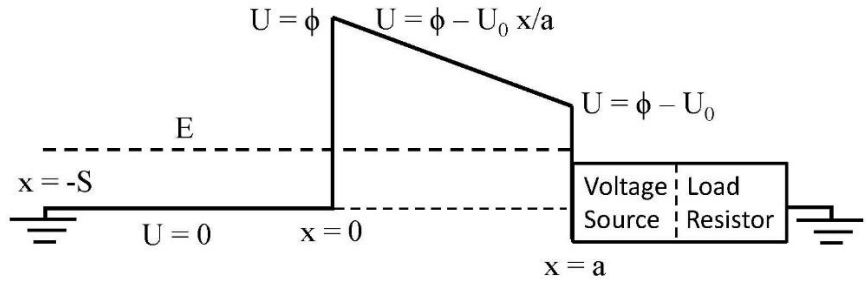

Fig. 2. Model for a closed-loop solution using a tunneling junction with a load resistor.

## IV. CONSISTENT STATIC SOLUTION OF THE SCHRÖDINGER EQUATION

Figure 3 shows the potential energy in a model of a nanoscale circuit with quantum tunneling in a constant axial electric field. Region 3 is the battery, having a smooth variation in the potential, so that Airy functions are used to model both the barrier and the battery. We follow a procedure similar to the derivation in Section 1 of the appendix. The symbols "U" and "V" are used separately for the potential energy and the voltage.

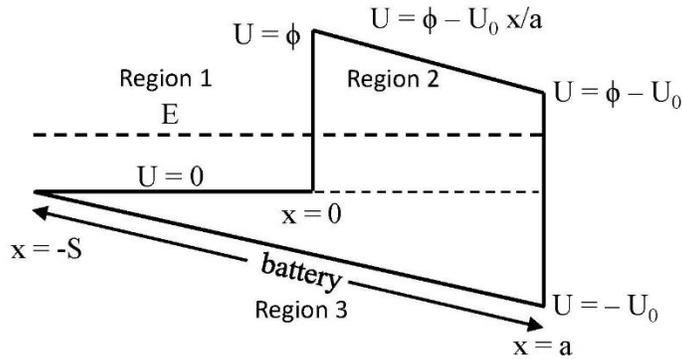

Fig. 3. Potential energy for a model of quantum tunneling in a static potential barrier.

In Region 3, representing the battery where -S < x < a, the potential is given by Eq. (2). We substitute this into the Schrödinger equation, in Eq. (3), to obtain Eq. (4).

$$U_3(x) = -U_0 \frac{(S+x)}{(S+a)} \tag{2}$$

$$\frac{\hbar^2}{2m} \frac{d^2\psi}{dx^2} + [E - U(x)]\psi = 0 \tag{3}$$

$$\frac{\hbar^2}{2m} \frac{d^2\psi_3}{dx^2} + \left[E + \frac{U_0 S}{(S+a)} + \frac{U_0 x}{(S+a)}\right]\psi_3 = 0 \tag{4}$$

A change of variables shown in Eq. (5) is used with Eq. (4) to obtain Eq. (6) where the coefficients $A_3$ and $B_3$ have units of meters. Then Eq. (6) is rearranged to obtain Eq. (7).

$$x = A_3 \xi + B_3 \tag{5}$$

$$\frac{\hbar^2}{2mA_3^2} \frac{d^2\psi_3}{d\xi^2} + \left[E + \frac{U_0 S}{(S+a)} + \frac{U_0(A_3\xi + B_3)}{(S+a)}\right]\psi_3 = 0 \tag{6}$$

$$\frac{d^2\psi_3}{d\xi^2} + \frac{2mA^2}{\hbar^2}\left[E + \frac{U_0(S+B_3)}{(S+a)}\right]\psi_3 + \frac{2mA_3^3 U_0}{\hbar^2(S+a)}\xi\psi_3 = 0 \tag{7}$$

Parameter $A_3$ is chosen by setting the coefficient of $\xi\psi_3$ in Eq. (7) to unity and parameter $B_3$ is chosen so that the quantity in brackets in Eq. (7) is zero. Thus, parameters $A_3$ and $B_3$ are given in Eqs. (8) and (9) and Eq. (6) is simplified to give Eq. (10). Note that $A_3$ is greater than zero and $B_3$ is less than zero.

$$A_3 = \left[\frac{\hbar^2(S+a)}{2mU_0}\right]^{\frac{1}{3}} \tag{8}$$

$$B_3 = -(S+a)\frac{E}{U_0} - S \tag{9}$$

$$\frac{d^2\psi_3}{d\xi^2} + \xi\psi_3 = 0 \tag{10}$$

The solution of Eq. (10) is given in Eq. (11) where Ai and Bi are Airy functions [17]. Equation (5) is used with Eq. (11) to obtain Eq. (12) with x as the independent variable. Note that in Region 3, the sign of the argument for the Airy functions is negative to give a quasi-sinusoidal behavior because the energy is greater than the potential. However, in Region 2, depending on the value for the energy, it is possible to have tunneling through all or only a part of the length of the barrier. Taking the derivative of Eq. (12) gives Eq. (13) for the derivative of the wavefunction in Region 3.

$$\psi_3(\xi) = C_5 Ai(-\xi) + C_6 Bi(-\xi) \tag{11}$$

$$\psi_3(x) = C_5 Ai\left(\frac{B_3 - x}{A_3}\right) + C_6 Bi\left(\frac{B_3 - x}{A_3}\right) \tag{12}$$

$$\frac{d\psi_3}{dx} = -\frac{C_5}{A_3} Ai'\left(\frac{B_3 - x}{A_3}\right) - \frac{C_6}{A_3} Bi'\left(\frac{B_3 - x}{A_3}\right) \tag{13}$$

The wavefunction and its derivative in Region 1 where $-S < x < 0$ are given by Eqs. (14), (15), and (16).

$$\psi_1(x) = C_1 e^{-ik_1 x} + C_2 e^{ik_1 x} \tag{14}$$

$$k_1 = \frac{\sqrt{2mE}}{\hbar} \tag{15}$$

$$\frac{d\psi_1}{dx} = -ik_1 C_1 e^{-ik_1 x} + ik_1 C_2 e^{ik_1 x} \tag{16}$$

Following the procedure that was used in Region 3, the wavefunction and its derivative in Region 2 where $0 < x < a$ are given by Eqs. (17), (18), (19), and (20).

$$\psi_2(x) = C_3 Ai\left(\frac{B_2 - x}{A_2}\right) + C_4 Bi\left(\frac{B_2 - x}{A_2}\right) \tag{17}$$

$$A_2 = \left[\frac{\hbar^2 a}{2mU_0}\right]^{\frac{1}{3}} \tag{18}$$

$$B_2 = \frac{(\phi - E)}{U_0} a \tag{19}$$

$$\frac{d\psi_2}{dx} = -\frac{C_3}{A_2} Ai'\left(\frac{B_2 - x}{A_2}\right) - \frac{C_4}{A_2} Bi'\left(\frac{B_2 - x}{A_2}\right) \tag{20}$$

Next the two boundary conditions, that the wavefunction and its derivative are continuous, are applied at each of the three boundaries. First Eqs. (21) and (22) are obtained at $x = 0$ between Region 1 and Region 2.

$$C_1 + C_2 - C_3 Ai\left(\frac{B_2}{A_2}\right) - C_4 Bi\left(\frac{B_2}{A_2}\right) = 0 \tag{21}$$

$$ik_1 C_1 - ik_1 C_2 - \frac{C_3}{A_2} Ai'\left(\frac{B_2}{A_2}\right) - \frac{C_4}{A_2} Bi'\left(\frac{B_2}{A_2}\right) = 0 \tag{22}$$

Applying the boundary conditions at $x = a$, between Region 2 and Region 3 gives Eqs. (23) and (24).

$$C_3 Ai\left(\frac{B_2 - a}{A_2}\right) + C_4 Bi\left(\frac{B_2 - a}{A_2}\right) - C_5 Ai\left(\frac{B_3 - a}{A_3}\right) - C_6 Bi\left(\frac{B_3 - a}{A_3}\right) = 0 \tag{23}$$

$$\frac{C_3}{A_2} Ai'\left(\frac{B_2 - a}{A_2}\right) + \frac{C_4}{A_2} Bi'\left(\frac{B_2 - a}{A_2}\right) - \frac{C_5}{A_3} Ai'\left(\frac{B_3 - a}{A_3}\right) - \frac{C_6}{A_3} Bi'\left(\frac{B_3 - a}{A_3}\right) = 0 \tag{24}$$

Applying the boundary conditions at $x = -S$, between Region 3 and Region 1 gives Eqs. (25) and (26).

$$C_1 e^{ik_1 S} + C_2 e^{-ik_1 S} - C_5 Ai\left(\frac{B_3 + S}{A_3}\right) - C_6 Bi\left(\frac{B_3 + S}{A_3}\right) = 0 \tag{25}$$

$$ik_1 C_1 e^{ik_1 S} - ik_1 C_2 e^{-ik_1 S} - \frac{C_5}{A_3} Ai'\left(\frac{B_3 + S}{A_3}\right) - \frac{C_6}{A_3} Bi'\left(\frac{B_3 + S}{A_3}\right) = 0 \quad (26)$$

Equations (21) through (26) form a system of six simultaneous homogeneous equations in the six unknown coefficients. We define the elements of the matrix for this system as $M_{IJ}$ where the indices I and J, for the row and column numbers, each run from 1 through 6. For this system to have a unique solution the determinant must be zero. The determinant is labeled as "det" in Eq. (27) where there are only three non-zero products.

$$\det = M_{11} M_{22} M_{33} M_{44} M_{55} M_{66} + M_{12} M_{23} M_{34} M_{45} M_{56} M_{61} + M_{13} M_{24} M_{35} M_{46} M_{51} M_{62} = 0 \quad (27)$$

Substituting the expressions for the matrix elements into Eq. (3) and removing a common prefactor gives Eq. (28).

$$Ai\left(\frac{B_2 - a}{A_2}\right) Bi'\left(\frac{B_2 - a}{A_2}\right) Ai\left(\frac{B_3 + S}{A_3}\right) Bi'\left(\frac{B_3 + S}{A_3}\right) + Ai\left(\frac{B_2}{A_2}\right) Bi'\left(\frac{B_2}{A_2}\right) Ai\left(\frac{B_3 - a}{A_3}\right) Bi'\left(\frac{B_3 - a}{A_3}\right)$$
$$+ Ai'\left(\frac{B_2}{A_2}\right) Bi\left(\frac{B_2 - a}{A_2}\right) Ai'\left(\frac{B_3 - a}{A_3}\right) Bi\left(\frac{B_3 + S}{A_3}\right) e^{ik_1 S} = 0 \quad (28)$$

Note that the first two terms in Eq. (28) are real and the third term is complex because of the imaginary exponential. Thus, we require that Eq. (29) be satisfied so that there will be no imaginary component. Thus, there are two solutions given by Eqs. (30) and (31).

$$k_1 S = n\pi \quad (29)$$

For n even:

$$Ai\left(\frac{B_2 - a}{A_2}\right) Bi'\left(\frac{B_2 - a}{A_2}\right) Ai\left(\frac{B_3 + S}{A_3}\right) Bi'\left(\frac{B_3 + S}{A_3}\right) + Ai\left(\frac{B_2}{A_2}\right) Bi'\left(\frac{B_2}{A_2}\right) Ai\left(\frac{B_3 - a}{A_3}\right) Bi'\left(\frac{B_3 - a}{A_3}\right)$$
$$+ Ai'\left(\frac{B_2}{A_2}\right) Bi\left(\frac{B_2 - a}{A_2}\right) Ai'\left(\frac{B_3 - a}{A_3}\right) Bi\left(\frac{B_3 + S}{A_3}\right) = 0 \quad (30)$$

For n odd:

$$Ai\left(\frac{B_2 - a}{A_2}\right) Bi'\left(\frac{B_2 - a}{A_2}\right) Ai\left(\frac{B_3 + S}{A_3}\right) Bi'\left(\frac{B_3 + S}{A_3}\right) + Ai\left(\frac{B_2}{A_2}\right) Bi'\left(\frac{B_2}{A_2}\right) Ai\left(\frac{B_3 - a}{A_3}\right) Bi'\left(\frac{B_3 - a}{A_3}\right)$$
$$- Ai'\left(\frac{B_2}{A_2}\right) Bi\left(\frac{B_2 - a}{A_2}\right) Ai'\left(\frac{B_3 - a}{A_3}\right) Bi\left(\frac{B_3 + S}{A_3}\right) = 0 \quad (31)$$

Equation (29) requires that the length S of the pre-barrier region be an integer multiple of one-half of the De Broglie wavelength as shown in Eq. (32), where the De Broglie wavelength is defined in Eq. (33) with the symbol "p" standing for the momentum of the electron.

$$S = \frac{n \lambda_{dB}}{2} \quad (32)$$

$$\lambda_{dB} \equiv \frac{h}{p} \quad (33)$$

Standing waves must have a constant potential over a defined spatial extent so we do not see them in sections of our models that represent either an ideal resistor or an ideal battery.

# V. ALGORITHMS FOR APPLYING THE ANALYSIS IN SECTION IV.

**Algorithm 1.**

In the first algorithm we hold a, $V_0$, and n constant and vary E (thus varying S) to obtain a determinant of zero so that there may be a unique solution for the system of four simultaneous equations. The procedure for this algorithm is as follows:

1. Specify the fundamental constants: Planck's constant and the rest mass and charge of the electron.
2. Specify $\phi$, the work function for the cathode.
3, Specify a, the length of the tunneling junction.
4. Specify $V_0$, the applied potential.
5. Specify the integer n.
6. Specify a trial value for the energy E of the electron relative to the fermi level.
7. Calculate $k_1$, the propagation constant in the pre-barrier region using Eq. (15).
8. Calculate S, the pre-barrier length, using Eq. (29).
9. Calculate the parameters $A_3$, $B_3$, $A_2$, and $B_2$ for determining the Airy functions by using Eqs. (8), (9), (18), and (19).
10. Calculate the 12 Airy functions $Ai(B_2/A_2)$, $Ai'(B_2/A_2)$, $Bi'(B_2/A_2)$, $Ai[(B_2-a)/A_2]$, $Bi[(B_2-a)/A_2]$, $Bi'[(B_2-a)/A_2]$, $Ai[(B_3-a)/A_3]$, $Bi[(B_3-a)/A_3]$, $Bi'[(B_3-a)/A_3]$, $Ai[(B_3+S)/A_2]$, $Bi[(B_3+S)/A_2]$, and $Bi'[(B_3+S)/A_2]$.
11. If the integer n is even use Eq. (30) to obtain the error and if n is odd use Eq. (31).
12. Return to step 6 with another trial value for the energy E, preferably using iterations with the numerical method of Bisection until the solution converges to the eigenvalue for E at which the determinant is zero for the specified values of a and $V_0$, and n.

**Algorithm 2.**

In the second algorithm we hold E and n (and thus S), as well as $V_0$, constant and vary a to obtain a determinant of zero so that there may be a unique solution for the system of four simultaneous equations. This corresponds to scanning tunneling microscopy in the constant-potential mode. The procedure for this algorithm is as follows:

1. Specify the fundamental constants: Planck's constant and the rest mass and charge of the electron.
2. Specify $\phi$, the work function for the cathode.
3. Specify $V_0$, the applied potential.
4. Specify the integer n.
5. Specify the energy E of the electron relative to the fermi level.
6. Specify a trial value for a, the length of the tunneling junction.
7. Calculate $k_1$, the propagation constant in the pre-barrier region using Eq. (15).
8. Calculate S, the pre-barrier length, using Eq. (29).
9. Calculate the parameters $A_3$, $B_3$, $A_2$, and $B_2$ for determining the Airy functions by using Eqs. (8), (9), (18), and (19).
10. Calculate the 12 Airy functions $Ai(B_2/A_2)$, $Ai'(B_2/A_2)$, $Bi'(B_2/A_2)$, $Ai[(B_2-a)/A_2]$, $Bi[(B_2-a)/A_2]$, $Bi'[(B_2-a)/A_2]$, $Ai[(B_3-a)/A_3]$, $Bi[(B_3-a)/A_3]$, $Bi'[(B_3-a)/A_3]$, $Ai[(B_3+S)/A_2]$, $Bi[(B_3+S)/A_2]$, and $Bi'[(B_3+S)/A_2]$.
11. If the integer n is even use Eq. (30) to obtain the error and if n is odd use Eq. (31).

12. Return to step 6 with another trial value for a, the length of the tunneling junction, preferably using the method of Bisection in iterations, until the solution converges to the eigenvalue for the length a at which the determinant is zero for the specified values of $V_0$. n, and E.

**Algorithm 3.**

In the third algorithm we hold E and n (and thus S), and a constant and Vary $V_0$ to obtain a determinant of zero so that there may be a unique solution for the system of four simultaneous equations. The procedure for this algorithm is as follows:
1. Specify the fundamental constants: Planck's constant and the rest mass and charge of the electron.
2. Specify ϕ, the work function for the cathode.
3, Specify a, the length of the tunneling junction.
4. Specify the integer n.
5. Specify the energy E of the electron relative to the fermi level.
6. Specify a trial value for $V_0$, the applied potential.
7. Calculate $k_1$, the propagation constant in the pre-barrier region using Eq. (15).
8. Calculate S, the pre-barrier length, using Eq. (29).
9. Calculate the parameters $A_3$, $B_3$, $A_2$, and $B_2$ for determining the Airy functions by using Eqs. (8), (9), (18), and (19).
10. Calculate the 12 Airy functions $Ai(B_2/A_2)$, $Ai'(B_2/A_2)$, $Bi'(B_2/A_2)$, $Ai[(B_2-a)/A_2]$, $Bi[(B_2-a)/A_2]$, $Bi'[(B_2-a)/A_2]$, $Ai[(B_3-a)/A_3]$, $Bi[(B_3-a)/A_3]$, $Bi'[(B_3-a)/A_3]$, $Ai[(B_3+S)/A_2]$, $Bi[(B_3+S)/A_2]$, and $Bi'[(B_3+S)/A_2]$.
11. If the integer n is even use Eq. (30) to obtain the error and if n is odd use Eq. (31).
12. Return to step 6 with another trial value for the potential $V_0$, preferably using the method of Bisection in iterations, until the solution converges to the eigenvalue for the potential $V_0$ at which the determinant is zero for the specified values of a, n, and E.

Note that a single program may be used to implement all 3 of these algorithms because a, $V_0$, n and E may be entered independently.

## VI. EXAMPLES USING THE THREE ALGORITHMS PRESENTED IN SECTION V.

All three algorithms use the same parameters but these algorithms differ in choosing which parameters must be given and which are to be determined. In each case the solution is determined by changing the unknown parameter until the determinant is zero using Eqs. (30) or (31). Due to the unusually sharp dependence of the determinant on the parameters, we present the results of our simulations in tables with no figures. For example, graphs of the data in the first two tables would only show the sharp variation of the determinant by many orders of magnitude and would fail to show the unusually high precision with which the dependent variable is obtained where the determinant is zero. Graphs of the data for the third table would show only the approximately linear dependence of the required voltage $V_0$ on the length of the barrier a, as well as the length of the pre-barrier region S, without showing the deviation from this apparent linearity—especially at small values of the length a.

Table I shows the applied voltage $V_0$ that causes the determinant to be zero for a determined system of equations when the length of the barrier a is 0.5 nm and the integer n is 1. These calculations determine the potential $V_0$ for three eigenvalues of the energy; 0.1, 0.2, 0.3, and 0.4 eV. Please note that for these four different energies the value of $V_0$ is calculated to an average precision of 2.5 parts-per-million. Clearly Table I also shows that the value of the voltage at which the determinant is zero, which we call $V_{0I}$, for I = 1 to 4, decreases as the corresponding eigenvalue for the energy $E_I$ increases from 0.1 to 0.4 eV.

A least-squares linear regression of these data using the 4 data points for $V_{0I}$ and $E_I$ with the hypothesis that $V_{0I} = A + BE_I$ gives the result A = 10.47323 Volts with B = -2.30292 which has a mean absolute fractional error of 86 parts-per-million with this data set.

Table I. Values of the applied voltage $V_0$ required for a = 0.5 nm and n = 1 at four eigenvalues of the energy

| E = 0.1 eV | | | E = 0.2 eV | | | E = 0.3 eV | | | E = 0.4 eV | | |
|---|---|---|---|---|---|---|---|---|---|---|---|
| $V_0$ | Deter | $V_0$-mean μV | $V_0$ | Deter | $V_0$-mean μV | $V_0$ | Deter | $V_0$-mean μV | $V_0$ | Deter | $V_0$-mean μV |
| 10.2437800 | -80532 | -5.0 | 10.0108 | -524694 | -1000 | 9.7814881 | -499 | -6.0 | 9.55288460 | -707 | -6.0 |
| 10.2437805 | -72423 | -4.5 | 10.0109 | -471616 | -900 | 9.7814886 | -458 | -5.5 | 9.55289060 | -563 | -5.5 |
| 10.2437810 | -64314 | -4.0 | 10.0110 | -418538 | -800 | 9.7814901 | -333 | -4.0 | 9.55289610 | -431 | -4.0 |
| 10.2437815 | -56204 | -3.5 | 10.0111 | -365458 | -700 | 9.7814906 | -292 | -3.5 | 9.55290010 | -335 | -3.5 |
| 10.2437820 | -48095 | -3.0 | 10.0112 | -312379 | -600 | 9.7814911 | -251 | -3.0 | 9.55290360 | -251 | -3.0 |
| 10.2437825 | -39986 | -2.5 | 10.0113 | -259299 | -500 | 9.7814916 | -209 | -2.5 | 9.55290660 | -179 | -2.5 |
| 10.2437830 | -31876 | -2.0 | 10.0114 | -206219 | -400 | 9.7814921 | -168 | -2.0 | 9.55290910 | -119 | -2.0 |
| 10.2437835 | -23767 | -1.5 | 10.0115 | -153138 | -300 | 9.7814926 | -126 | -1.5 | 9.55291110 | -71 | -1.5 |
| 10.2437840 | -15657 | -1.0 | 10.0116 | -100057 | -200 | 9.7814931 | -85 | -1.0 | 9.55291260 | -35 | -1.0 |
| 10.2437845 | -7548 | -0.5 | 10.0117 | -46976 | -100 | 9.7814936 | -44 | -0.5 | 9.55291360 | -11 | -0.5 |
| 10.2437850 | [0] | [0] | 10.0118 | [0] | [0] | 9.7814941 | [0] | [0] | 9.55291410 | [0] | [0] |
| 10.2437855 | 8671 | 0.5 | 10.0119 | 59188 | 100 | 9.7814946 | 39 | 0.5 | 9.55291460 | 13 | 0.5 |
| 10.2437860 | 16780 | 1.0 | 10.0120 | 112271 | 200 | 9.7814951 | 81 | 1.0 | 9.55291560 | 37 | 1.0 |
| 10.2437865 | 24889 | 1.5 | 10.0121 | 165354 | 300 | 9.7814956 | 122 | 1.5 | 9.55291710 | 73 | 1.5 |
| 10.2437870 | 32999 | 2.0 | 10.0122 | 218437 | 400 | 9.7814961 | 163 | 2.0 | 9.55291910 | 121 | 2.0 |
| 10.2437875 | 41108 | 2.5 | 10.0123 | 271521 | 500 | 9.7814966 | 205 | 2.5 | 9.55292160 | 181 | 2.5 |
| 10.2437880 | 49218 | 3.0 | 10.0124 | 324605 | 600 | 9.7814971 | 246 | 3.0 | 9.55292460 | 253 | 3.0 |
| 10.2437885 | 57327 | 3.5 | 10.0125 | 377690 | 700 | 9.7814976 | 288 | 3.5 | 9.55292810 | 337 | 3.5 |
| 10.2437890 | 65436 | 4.0 | 10.0126 | 430775 | 800 | 9.7814981 | 329 | 4.0 | 9.55293210 | 433 | 4.0 |
| 10.2437895 | 73546 | 4.5 | 10.0127 | 483860 | 900 | 9.7814986 | 370 | 4.5 | 9.55293660 | 541 | 4.5 |
| 10.2437900 | 81655 | 5.0 | 10.0128 | 536946 | 1000 | 9.7814991 | 412 | 5.0 | 9.55294160 | 661 | 5.0 |

Table II shows the effect of using the integer n greater than 1. This is accomplished in the first instance by comparing the value of the applied voltage $V_0$ required to obtain the eigenvalue E = 0.2 eV with a barrier length a of 0.5 nm using both n = 1 and 2. Note that the voltage steps which are required to obtain the same change in the determinant are reduced by a factor of approximately 1,000 when going from n = 1 to n = 2. We have not yet made studies using n greater than 1 because of the high precision that would be required for corresponding measurements. Note that the pre-barrier length S must be increased in proportion to n as shown in Eq. (32).

Table II. Values of the applied voltage $V_0$ required for a = 0.5 nm and the eigenvalue E = 0.2 eV at two values of n
S = 1.371 nm for n = 1 and 2.742 nm for n = 2.

| n = 1 | | $V_0$-mean | n = 2 | | $V_0$-mean |
| $V_0$ | Deter | µV | $V_0$ | Deter | µV |
|---|---|---|---|---|---|
| 10.0108 | -524694 | -1000 | 10.01186801 | -767680 | -1.10 |
| 10.0109 | -471616 | -900 | 10.01186812 | -690957 | -0.99 |
| 10.0110 | -418538 | -800 | 10.01186823 | -614233 | -0.88 |
| 10.0111 | -365458 | -700 | 10.01186834 | -537510 | -0.77 |
| 10.0112 | -312379 | -600 | 10.01186845 | -460786 | -0.66 |
| 10.0113 | -259299 | -500 | 10.01186856 | -384063 | -0.55 |
| 10.0114 | -206219 | -400 | 10.01186867 | -307339 | -0.44 |
| 10.0115 | -153138 | -300 | 10.01186878 | -230616 | -0.33 |
| 10.0116 | -100057 | -200 | 10.01186889 | -153892 | -0.22 |
| 10.0117 | -46976 | -100 | 10.01186900 | -77169 | -0.11 |
| 10.0118 | [0] | [0] | 10.01186911 | (0) | (0) |
| 10.0119 | 59188 | 100 | 10.01186922 | 76278 | 0.11 |
| 10.0120 | 112271 | 200 | 10.01186933 | 153000 | 0.22 |
| 10.0121 | 165354 | 300 | 10.01186944 | 229730 | 0.33 |
| 10.0122 | 218437 | 400 | 10.01186955 | 306450 | 0.44 |
| 10.0123 | 271521 | 500 | 10.01186966 | 383170 | 0.55 |
| 10.0124 | 324605 | 600 | 10.01186977 | 459900 | 0.66 |
| 10.0125 | 377690 | 700 | 10.01186988 | 536620 | 0.77 |
| 10.0126 | 430775 | 800 | 10.01186999 | 613340 | 0.88 |
| 10.0127 | 483860 | 900 | 10.01187010 | 690070 | 0.99 |
| 10.0128 | 536946 | 1000 | 10.01187021 | 766790 | 1.10 |

Table III shows the values of the applied voltage that is required for the determinant to be zero at different values of the distance a for three eigenvalues of the energy. While the voltage is approximately proportional to the distance, note that the magnitude of the electric field, which is constant within the barrier, becomes smaller as the distance a is reduced, and the electric field is substantially increased for a equal to 0.10 nm.

Table III. Values of the applied voltage $V_0$ required for different values of the distance a with n = 1. The specified values of S are required at each energy.

| a nm | E = 0.10 eV S = 1.939 nm | | E = 0.20 eV S = 1.371 nm | | E = 0.30 eV S = 1.120 nm | |
|---|---|---|---|---|---|---|
| | $V_0$ | $E_x$ V/nm | $V_0$ | $E_x$ V/nm | $V_0$ | $E_x$ V/nm |
| 0.50 | 10.243785 | 20.48757 | 10.011800 | 20.02360 | 9.781494 | 19.56299 |
| 0.45 | 9.219398 | 20.48755 | 9.010553 | 20.02345 | 8.803167 | 19.56259 |
| 0.40 | 8.195007 | 20.48752 | 8.009274 | 20.02319 | 7.824711 | 19.56178 |
| 0.35 | 7.170608 | 20.48745 | 7.007925 | 20.02264 | 6.845987 | 19.55996 |
| 0.30 | 6.146192 | 20.48731 | 6.006395 | 20.02132 | 5.866630 | 19.55543 |
| 0.25 | 5.121733 | 20.48693 | 5.004425 | 20.01770 | 4.885502 | 19.54201 |
| 0.20 | 4.097158 | 20.48579 | 4.001046 | 20.00523 | 3.898040 | 19.49020 |
| 0.15 | 3.072143 | 20.48095 | 2.991268 | 19.94179 | 2.868900 | 19.12600 |
| 0.10 | 2.044400 | 20.44400 | 1.871302 | 18.71302 | (no root) | |

## VII. PROCEDURE FOR DETERMINING EACH EIGENSOLUTION

First, we change the set of six equations, Eqs. (21), (22), (23), (24), (25) and (26) to form a set where all of the coefficients are divided by the coefficient $C_1$, to normalize them relative to the incident wave in the pre-barrier region. Then the terms that do not contain these ratios are moved to the RHS to create a new set of equations. This gives the following system of six equations with the five normalized coefficients that are labeled as $X_I$ which is defined as $C_I/C_1$ for I = 2, 3, 4, 5, 6 in Eqs. (34), (35), (36), (37), (38), and (39).

$$X_2 - Ai\left(\frac{B_2}{A_2}\right)X_3 - Bi\left(\frac{B_2}{A_2}\right)X_4 = -1 \qquad (34)$$

$$ik_1 X_2 + \frac{1}{A_2} Ai'\left(\frac{B_2}{A_2}\right)X_3 + \frac{1}{A_2} Bi'\left(\frac{B_2}{A_2}\right)X_4 = ik_1 \qquad (35)$$

$$Ai\left(\frac{B_2 - a}{A_2}\right)X_3 + Bi\left(\frac{B_2 - a}{A_2}\right)X_4 - Ai\left(\frac{B_3 - a}{A_3}\right)X_5 - Bi\left(\frac{B_3 - a}{A_3}\right)X_6 = 0 \qquad (36)$$

$$\frac{1}{A_2} Ai'\left(\frac{B_2 - a}{A_2}\right)X_3 + \frac{1}{A_2} Bi'\left(\frac{B_2 - a}{A_2}\right)X_4 - \frac{1}{A_3} Ai'\left(\frac{B_3 - a}{A_3}\right)X_5 - \frac{1}{A_3} Bi'\left(\frac{B_3 - a}{A_3}\right)X_6 = 0 \qquad (37)$$

$$e^{-ik_1 S} X_2 - Ai\left(\frac{B_3 + S}{A_3}\right)X_5 - Bi\left(\frac{B_3 + S}{A_3}\right)X_6 = -e^{ik_1 S} \qquad (38)$$

$$ik_1 e^{-ik_1 S} X_2 + \frac{1}{A_3} Ai'\left(\frac{B_3 + S}{A_3}\right)X_5 + \frac{1}{A_3} Bi'\left(\frac{B_3 + S}{A_3}\right)X_6 = ik_1 e^{ik_1 S} \qquad (39)$$

In Section VI we presented three algorithms where in each one three members of the set a, n, $V_0$, and E are specified and the remaining member is determined by iteration. In Section VII we presented examples using these algorithms with different values of a, n, $V_0$, and E. Any of these sets of a, n, $V_0$, and E, or others for which the determinant is also zero, may be used with the new system of Eqs. (34) to (39) to determine the five normalized coefficients as the corresponding eigensolutions. However, note that this system of six equations in five unknowns is overdetermined. Thus, with no loss of generality, we may solve a set of any five of these equations to determine the five normalized coefficients for the corresponding eigensolutions. Our remaining task is to make the calculations to obtain such complete solutions.

**VIII. EXTENSION TO TIME-DEPENDENT APPLICATIONS**

The mechanism studied by Tien and Gordon [18] requires photon processes with superconducting electrodes to provide a time-dependent tunneling current and their solution of the time-dependent Schrödinger equation is not unique. A separate exact solution of the time-dependent Schrödinger equation that is also based on photon processes was presented by Zhang and Lau [19].

We have generated microwave frequency combs by focusing a mode-locked ultrafast laser on the tunneling junction of a scanning tunneling microscope [5] and our analysis of these measurements suggests that quantum processes are not involved in this application [20]. However, others have described photon-assisted processes in analyses [18], [21] and measurements [22], [23], [24]. Their solutions are strongly dependent on the gap length and would not be expected under the conditions for our measurements [5], [7], [25], [26].

In our first measurements of time-dependent quantum tunneling we connected the primary windings of three audio-frequency transformers in series with a vacuum field emission tube and a DC high-voltage power supply. Two audio-frequency oscillators and an oscilloscope were connected to the secondary windings of the three transformers. We measured tunneling currents at the frequencies of the two oscillators as well as harmonics and other components at the mixer frequencies. Quasistatic solutions, obtained by substituting an expression for the time-dependent voltage into the Fowler-Nordheim equation for field emission, were shown to be consistent with our measurements. The photon energy in these measurements was typically 1 peV so it would not be reasonable to attribute these results to photon processes.

To further justify the use of quasistatic approximations it is necessary to consider the ratio of the length of the tunneling junction to the laser wavelength in our measurements of laser-assisted scanning tunneling microscopy [5]. The center wavelength for the Ti: sapphire laser that we used to generate the microwave frequency comb is 800 nm, so a 0.4 nm tunneling junction has a length that is only 0.05 percent of the wavelength. Thus, we feel justified to make the approximation of substituting the time-dependent potential of $U_0 + U_1 \cos(\omega t)$ in which there is both DC and a sinusoid, or simply $U_1 \cos(\omega t)$, or other expressions such as that for a mode-locked laser [20] in place of the $U_0$ in the static solutions that are considered in this paper.

Figure 4 is an example of an equivalent circuit that could be used to extend the static solutions by the quasi-static approximation. Note that the time-dependent tunneling current is divided between the load resistance and the shunting capacitance of the tunneling junction to provide some cut-off at extremely high frequencies. We measure the microwave frequency comb using a spectrum analyzer with a Bias-T in the sample circuit of an STM (UHV700, RHK Technology). The power at the first 200 harmonics (74.254 MHz to 14.85 GHz) varies inversely as the square of the frequency which we attribute to current division between the shunting capacitance of 6.4 pF associated with the tunneling junction and the 50-Ω network analyzer [5]. We would expect this shunting would be considerably reduced in a nanoscale circuit as part of a full-size instrument.

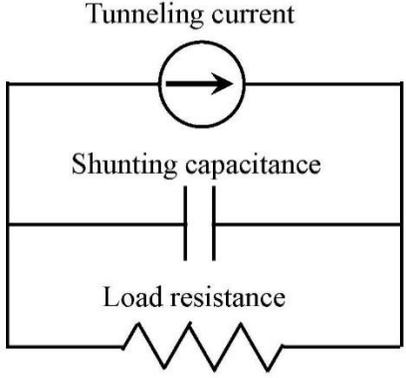

Fig. 4. Equivalent circuit for the quasi-static solutions.

## IX. CONCLUSIONS

We acknowledge that others have developed a variety of numerical methods to model nanoscale circuits [27], [28], [29], [30], [31], [32]. The work by Pierantoni, Mencarelli, and Rozzi [32] is especially pertinent to our work because they mention the possibility of ballistic transport within nanoscale devices. However, they do not show the significance of the mean-free path for electrons which has been determined for 20 different metals by Gall [14]. These three authors refer to measurements which show that the current in CNT FETs may be dramatically altered by small changes in the voltage at the gate electrode but they did not show a relationship between this behavior and the transport over long distances. Here we have defined a relatively simple analytical approach for determining the effects of the extended propagation of the electrons at specific eigenvalues of the energy.

We recognize that there are two applications for our new approach: (1) understanding and mitigating the effects which have already been mentioned by others with present nanoscale circuits [27], and (2) developing new devices with improved performance. The new devices could be self-contained such as by providing a voltage source using a nanoscale antenna driven by an external laser. It is also possible to have leads connecting to an external voltage source, or to monitor the voltage across a nanoscale load resistor. This would enable interfacing as part of a full-size instrument such as in scanning tunneling microscopy or scanning frequency comb microscopy [7]. We are addressing possible applications such as providing sub-nanometer resolution in the carrier profiling of semiconductors for metrology. Present methods are not adequate for this purpose which has led to a crisis in the semiconductor industry at and below the 7-nm technology node [7], [33]. Dr. Hagmann leads a SEMI Standards working group that is developing new metrology standards with presently available instrumentation to mitigate this problem.

**ACKNOWLEDGMENTS**

Dr. Hagmann is grateful to Rolf Landauer and Markus Büttiker who encouraged him to publish his initial work on the numerical modeling of laser-assisted quantum tunneling in 1995, and to Professor Marwan Mousa, at Mu'tah University in Jordan, who made it possible to make time-dependent measurements of laser-assisted field emission during his sabbatical visit to Florida International University from 1999-2000. We are also grateful to Dmitry Yarotski who made it possible to extend these measurements to laser-assisted Scanning Tunneling Microscopy in visits to the Center for Integrated Nanotechnologies (CINT) at Los Alamos National Laboratory from 2008 to 2017. This work has been sponsored by the National Science Foundation under Grant 1648811 and the U.S. Department of Energy under Award DE-SC000639.